\documentclass[10pt,prl,twocolumn,showpacs,amsmath,amssymb]{revtex4}

\usepackage{graphicx}
\usepackage{dcolumn} 
\usepackage{bm}      

\begin{document}

\title{Micron-scale droplet deposition on a hydrophobic surface using a retreating syringe}

\author{Bian Qian}
\author{Melissa Loureiro}
\author{David Gagnon}
\author{Anubhav Tripathi}
\author{Kenneth S. Breuer}
\email{kbreuer@brown.edu}
\affiliation{Division of Engineering, Box D, Brown University, Providence, RI
02912, USA}
\begin{abstract}
Droplet deposition onto a hydrophobic surface is studied experimentally and numerically.  A wide range of droplet sizes can result from the same syringe, depending strongly on the needle retraction speed.  Three regimes are identified according to the motion of the contact line.  In Region I, at slow retraction speeds, the contact line expands and large droplets can be achieved.  In Region II, at moderate needle speeds, a quasi-cylindrical liquid bridge forms resulting in drops approximately the size of the needle.  Finally, at high speeds (Region III), the contact line retracts and droplets much smaller than the syringe diameter are observed. Scaling arguments are presented identifying the dominant mechanisms in each regime.  Results from nonlinear numerical simulations agree well with the experiments, although the accuracy of the predictions is limited by inadequate models for the behavior of the dynamic contact angle.
\end{abstract}


\maketitle

Contact dispensing methods of fluids are widely used in a variety of applications including direct scanning probe lithography~\cite{ginger2003}, micromachined fountain-pen techniques~\citep{Deladi2004,moldovan2006},  electrowetting-assisted drop deposition~\cite{leichle2007} and biofluid dispensing applications~\cite{Ben2007,meister2004}.  The process is, at first glance, straightforward and is initiated by the formation of a liquid bridge between the substrate and a dispensing syringe. As the syringe retreats, the liquid bridge stretches, grows and breaks, leaving a drop on the substrate. A seemingly simple question can be asked - how does the drop size depend on the syringe geometry, speed and the fluid properties?  A comprehensive answer must consider the stability of the liquid bridge and the physics of the moving contact line at the liquid-air-solid interface - both difficult problems.  Theoretical studies of liquid bridge stability date back to Rayleigh~\cite{rayleigh1878}, and have been extended to include gravity and non-cylindrical geometries~\cite{slobozhanin1997,meseguer1984}. In addition, the nonlinear dynamics have been solved numerically, using both 2-D (axisymmetric) \cite{Liao2006} and 1-D (slender-jet)~\citep{eggers1994,zhang1996} models. Previous work has concentrated on geometries in which the contact line is \emph{pinned} at both ends of the liquid bridge~\cite{tripathi2000,zhang1996}, and there are only a few results that couple the liquid bridge with a moving contact line~\cite{amberg2007,Panditaratne_PhD2003}. A possible reason for this is the difficulty in solving the flow near the contact line where the continuum equations are invalid~\cite{huh1971,stone-eggers} and a microscopic description must be imposed (e.g.~\cite{bonn2008}). In this letter, we focus on the physics of drop dispensing on a flat, smooth, hydrophobic substrate in which the contact line is free to move and is inherently coupled with the liquid bridge stability. Experiments and numerical simulations are used to identify a range of complex flow phenomena which enable the deposited drop size to vary by two orders of magnitude as the syringe retraction speed is changed.

\begin{figure}
\includegraphics[width=\columnwidth]{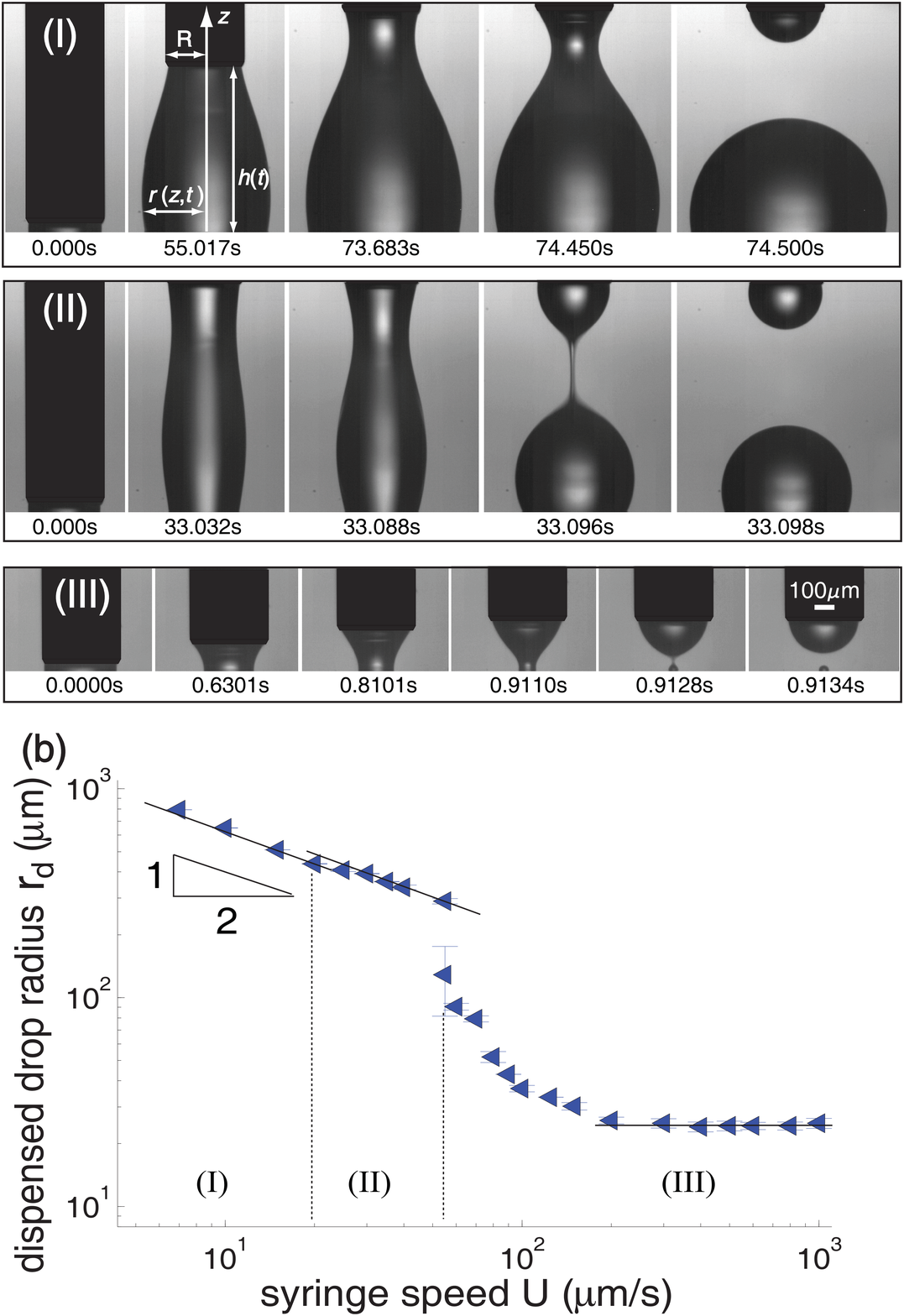}
\caption{(color online) Typical sequence of images of drop dispensing at syringe speeds of (I) 15\,$\mu m/s$, (II) 35\,$\mu m/s$ and (III) 400\,$\mu m/s$. The syringe radius is 205\,$\mu m$. (b) Dispensed drop radius, $r_d$ vs. syringe speed, $U$, illustrating the three regions: expanding contact line (I), pinned contact line (II) and retreating contact line (III).}
\label{seqimg} %
\end{figure}

In our experiment, a stainless steel syringe (typical radius, $R = 200\mu m$) is mounted vertically on a computer-controlled stage.  The syringe is connected by a small tube to a 10cc barrel mounted on the same stage. This configuration maintains a constant hydrostatic head, $H$, at the syringe tip ($H \sim 4$cm). The fluid (a 85-15 mixture by volume of glycerol and water) has viscosity $\mu = 84$\,{cP} and surface tension $\gamma = 0.063$\,{N/m}. The fluid exhibits a static contact angle of $\sim 90^\circ$ with the substrate,  a smooth glass slide coated with a monolayer of octadecyltrichlorosilane (OTS). The syringe is brought down towards the substrate, stopping $\sim 40 \mu m$ above the surface so that the meniscus touches the substrate and spreads, partially wetting the surface to form a stable drum-shaped liquid bridge (Fig.~\ref{seqimg}).  As the syringe retracts at a constant speed, $U$, the liquid bridge elongates and evolves due to the changing height, $h$, fluid flowing into the bridge through the syringe, (characterized by an inflow velocity, $u_f$, and the motion of the contact line between the bridge and the substrate (characterized by a contact line position, $r$, and speed, $u_c$).  At a critical height, $h_p$, the liquid bridge becomes unstable and pinches off rapidly, leaving a drop on the surface. A high speed camera (Photron APX) equipped with a 5X Mitutuyo lens was used to capture the motion of drop dispensing at frames rates up to 10kfps, with a resolution of $3.33\mu m/\mathrm{pixel}$.  The experiment was carried out using several syringe diameters, hydrostatic pressures and retraction speeds, and conducted multiple times to ensure repeatability.

\begin{figure}
\includegraphics[width=2.75in]{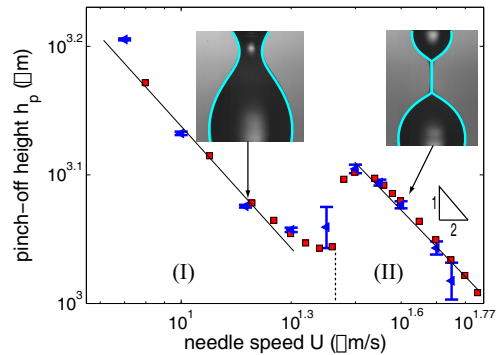}
\caption{(color online) Pinch-off height, $h_p$, in regions of expanding (I) and  fixed (II) contact line.  Numerical simulations ($\Box$) and experimental measurements ($\triangleleft$) are shown. For this case, $R = 205\mu m, Oh=0.087,Bo=0.0079$. The insets show measured and computed shapes of the liquid bridge prior to pinch-off (numerical solution: cyan line).}
\label{simuldrop}
\end{figure}

Given the simplicity of the experiment, the resulting drop size, $r_d$, shows a surprisingly complex dependence on the syringe speed (Fig.~\ref{seqimg}).  The data can be divided into three regions, categorized according the to motion of the contact line. At low retraction speeds (Fig.~\ref{seqimg}-I), the flow from the syringe into the liquid column is relatively high, and a bulging liquid bridge forms. The contact angle on the surface exceeds its equilibrium value, and the contact line expands outwards. In this regime, arbitrarily large drops can be formed, with the drop radius scaling with $U^{-1/2}$. As one increases the syringe speed (Fig.~\ref{seqimg}-II), the bridge elongation balances the incoming flow and the contact line becomes stationary.  In this regime, a quasi-cylindrical liquid bridge forms, finally pinching off as it becomes unstable. The resultant drop size still scales with $U^{-1/2}$. However, there is a jump in the drop size between the fixed and expanding contact line regimes due to a jump in the pinch-off height, $h_p$, (Fig.~\ref{simuldrop}). A third regime is achieved by increasing the syringe speed further (Fig.~\ref{seqimg}-III) The liquid bridge initially adopts a catenoidal shape, but as its contact angle falls below the equilibrium value, the contact line retreats \emph{inward} until, at some critical point, the liquid bridge become unstable and pinches off.
The rapid pinching motion drags the contact line inwards at a very high speed ($\sim 100$ times the syringe speed), and due to this rapid retreating motion, a small drop is deposited, with a diameter that continues to decrease until a minimum drop size is reached, independent of retraction speed, with a diameter approximately one tenth that of the dispensing syringe.

\begin{figure}
\includegraphics[width=2.75in]{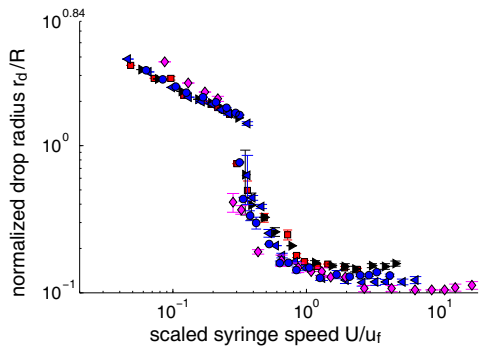}
\caption{(color online) Normalized drop sizes, $r_d/R$, at different scaled syringe speeds $U/u_f$ for three syringe radii: 205\,$\mu m$ ($\triangleleft$), 255\,$\mu m$ ($\bigcirc$) and 320\,$\mu m$ ($\Box$). The driving pressure, $\Delta p = \rho g H-({\gamma}/{R})$ is 172(\,{Pa}). Also shown are drop sizes obtained using an $R =  205 \mu m$ syringe and two effective driving pressures: $\Delta p  = 52.0$\,{Pa} ($\Diamond$) and 291\,{Pa}($\triangleright$). The syringe speed is scaled by the flow speed, $u_{f} = ({R^2}/{8\mu L})(\rho g H-{\gamma}/{R})$.}
\label{scale_dropsize} %
\end{figure}

We focus first on region II where the contact line is stationary, and the bridge near-cylindrical. In this region the syringe speed is much smaller than the capillary wave speed,  $U/\sqrt{\gamma/ \rho R} = {\cal O}(10^{-4})$, viscous and inertial forces are small, and the liquid bridge can be considered as quasi-static.  Since the Bond number,  $Bo\equiv g\rho R^2/\gamma$, is small  (${\cal O}(10^{-2})$), the hydrostatic pressure in the column is approximately constant, dominated by the Laplace pressure set by the needle curvature, $\gamma/R$.  This state is amenable to the the classical Rayleigh stability analysis, which predicts that the height-to-radius, $\Lambda = h_p/R$, must be less than $2\pi$ to maintain stability~\cite{rayleigh1878}.  Furthermore, for a marginally-stable liquid bridge (i.e. $\Lambda/2\pi = 1 + \epsilon$),  the dimensionless volume, $V\equiv v_p/\pi R^2 h_p$, is given by $V = 1 + 2\epsilon + 5 \epsilon^2/2+{\cal O}(\epsilon^3)$~\cite{slobozhanin1997}. Rearranging this equation gives $V= (\Lambda/2\pi)^2 + {\cal O}(\epsilon^2)$, from which we find that $v_p \sim h_p^3$. Assuming that the liquid flow from the syringe into the liquid bridge, $u_{f}$, is constant (driven by the net pressure difference, $\rho g H - \gamma/R$), both the volume and the height increase linearly with time: $v \sim u_f t; h \sim Ut$, from which it is easily derived that the pinch-off height at which the bridge becomes unstable, $h_p$, scales like $U^{-1/2}$, that the breaking time, $t_p$, scales like $U^{-3/2}$, and that the drop radius, $r_d$, scales like $U^{-1/2}$. The experiments support this scaling argument very well (Figs.~\ref{seqimg}-b,\ref{simuldrop}, and~\ref{scale_dropsize}).

In region I, where the contact line expands, the bridge is no longer cylindrical and the Rayleigh stability criteria can be modified by a small parameter, $\delta = (1-w^2)/(1+w^2)$, where $w=r(h)/r(0)$ is ratio of the upper and lower contact line radii~\cite{meseguer1984}. In agreement with the theory, we do see a decrease in the pinch-off height, $\Lambda$, just as the contact line begins to expand ($w < 1$), although $\delta$ quickly becomes too large for the perturbation analysis to remain valid. To address this, we employ a numerical model, previously used in studying jet breakup and the stretching of a pinned liquid bridge~\cite{eggers1994,zhang1996} with modified boundary conditions to include moving contact lines. The non-dimensionalized  equations for conservation of mass and momentum are given by:
\begin{equation}
       \partial_t{r} + ur'=-ru'/2, \label{eq_mass}
\end{equation}
\begin{equation}
       (\partial_t{u} + uu')=-\kappa '+\frac{3 Oh}{r^2}[(r^2u')']-Bo , \label{eq_momentum}
\end{equation}
while the evolution of the mean curvature, $\kappa$, (included to accurately predict the breakup beyond the validity of slender-jet approximation~\cite{eggers1994,yildirim2001}) is described by:
\begin{equation}
       \kappa=\frac{1}{r(1+r'^2)^{1/2}} - \frac{r''}{(1+r'^2)^{3/2}} . \label{eq_curvature}
\end{equation}
Here, $u(z,t)$ and  $r(z,t)$ are the axial flow speed and column radius, normalized by the capillary wave speed, $u_{cp}=\sqrt{\gamma/ \rho R}$ and the syringe radius, $R$ respectively. Time, $t$, is normalized by the capillary time, $\sqrt{\rho R^3/\gamma}$. A prime denotes the partial derivative with respect to the axial coordinate, $z$. For our experiments, Ohnesorge number $Oh \equiv \mu/\sqrt{\rho R \gamma} \approx {\cal O}(1)$.  Note that the Weber number, $We = \sqrt{ U/u_{cp} }$, should be small for the equations to remain valid~\cite{yildirim2001}.

\begin{figure}
\includegraphics[width=2.75in]{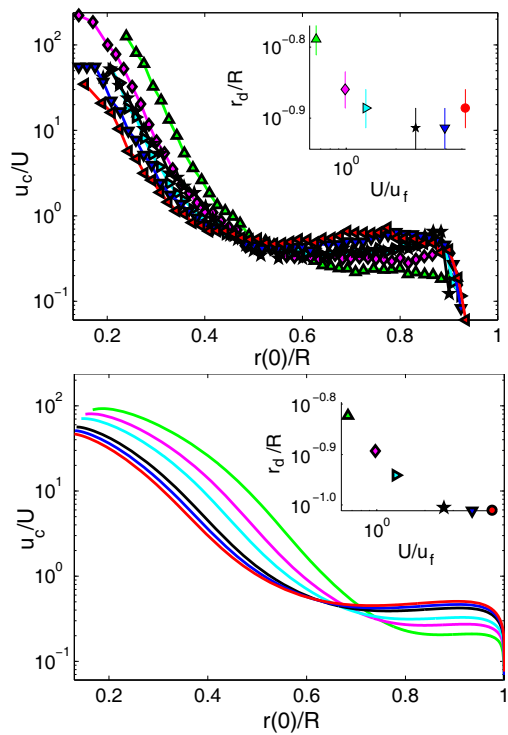}
\caption{(color online) (a) Contact line speed, $u_{c}/U$, vs. contact line location, $r(0,t)/R$, for various scaled syringe speeds: $U/u_{f}$ = 0.66($\bigtriangleup$), 0.99($\Diamond$), 1.31($\rhd$), 2.63($\bigstar$), 3.94($\bigtriangledown$) and 5.25($\bigcirc$). The inset shows the corresponding final drop size. (b) Corresponding numerical solution.} \label{contactline} %
\end{figure}

The boundary conditions require that at the top of the liquid bridge, $h=Ut$, and that the contact line is pinned, $r(h)=R$. The pressure here is the hydrostatic head minus the pressure drop due to the flow: $p(h) = \rho g H - 8\mu u_f L/R^2$ ($L$ is the length of the syringe), where the inflow velocity, $u_f$, is evaluated as $u_f = u(h)$. At the substrate, $z = 0$, a solid-wall boundary condition is imposed, $u(0)=0$. This last boundary condition is quite subtle, since the contact line may move with time, and this must be consistent with the solution of the model equations. For a fixed contact line, $r(0)$ is a constant and the apparent contact angle varies between $(\theta_{a},\theta_{r})$ - the advancing and receding equilibrium contact angles. For a moving contact line, the contact angle deviates from the equilibrium contact angle in order to balance the viscous drag.  In general this is a function of the contact line speed, $u_c = \partial_t r$. Predicting the contact line behavior and the dynamic contact angle is an area of active research (e.g. \cite{bonn2008}).  For simplicity, we use constant values for ($\theta_a, \theta_r$) of $100^\circ$ and $80^\circ$ respectively. The equations (\ref{eq_mass})-(\ref{eq_curvature}) were solved numerically \cite{eggers1994}.  Fig.~\ref{simuldrop} indicates excellent agreement achieved between the numerical solutions and the experimental measurements of the pinch-off height, bridge shape, and dependence on the syringe speed.

The speed at which liquid flows into the bridge, $u_{f}$, is a critical scaling velocity, affecting the pressure boundary condition at the top of the bridge and the rate at which the bridge volume grows (which in turn plays a central role in the bridge stability and subsequent pinch-off). We can estimate $u_{f}$ assuming Poiseuille flow through a syringe of length $L$, radius $R$, and driven by the net pressure difference, $\Delta p = \rho g H - \gamma/R$.  This estimate differs by only a few percent from the value predicted by the 1-D equations, The appropriateness of this scaling is confirmed in Fig.~\ref{scale_dropsize} which shows the resultant drop size $r_d/R$,  versus the scaled retraction speed, $U/u_{f}$, for a series of experiments obtained using three syringe sizes and three hydrostatic pressures (yielding values of $u_{f}$ that ranged from $46-414\mu m/s$).

The physics of the drop deposition changes abruptly in region III where the contact line begins to retreat.  High speed imaging was used to measure the contact line position $r(0,t)$ from which its speed, $u_c$, was calculated (Fig.~\ref{contactline}(a)). Initially, the contact line is at its maximum ($r(0)/R \approx 1$), and has zero speed. As the syringe begins to retreat, there is a short period of acceleration, after which time the contact line moves inward with approximately constant speed. However, at a critical radius, approximately $r(0)/R \sim 0.5$, we see a dramatic acceleration  with $u_c/U$ reaching ${\cal O}(100)$ immediately prior to pinch-off. In the constant speed region, $u_c/u_{cp}\ll 1$,  inertial and viscous forces are negligible and the contact line speed is thus determined solely from the mass balance and pressure equilibrium in the bridge. Applying this balance, and the fact that our estimate for $u_f$ does not depend on $U$, it is easy to show geometrically that $u_c/U \sim 1 - u_f/U$, a trend confirmed in both the experiments and simulations (Fig.~\ref{contactline}).

The high-speed contact line phase is driven by the pinch-off instability, during which time a strong capillary force pulls the contact line inward at increasing speed. As $u_c/u_{cp}$ approaches unity,  the viscous forces become significant, and the contact line acceleration decreases immediately prior to pinch-off. We also see that the radial location at which the pinch-off instability initiates moves inward as $U$ increases, and that this affects the final drop size (Fig~\ref{contactline}a-inset). A detailed stability analysis of the asymmetric liquid bridge with a moving contact line explains that the smaller bridge volume corresponding to the higher syringe speed postpones the pinch-off instability, resulting in a smaller drop \cite{qian2008}. However, as the syringe speed increases further, the location at which the contact line begins its rapid acceleration moves back out, and the drop size increases (Fig.~\ref{contactline}(a) inset). A possible reason for this may be that the dynamic contact angle decreases, which destabilizes the liquid bridge earlier \cite{qian2008}.  The numerical model yields good comparisons with the experimental results (Fig.~\ref{contactline}b), capturing the general behavior in the constant speed region (including the increase in $u_c/U$ with $U$) as well as the onset of rapid acceleration prior to pinch-off. However, the model predicts the critical radius to be larger than that seen in the experiment. We believe that the reason for this discrepancy is that the numerical simulation uses a static retreating contact angle, $\theta = 80^\circ$,  while the contact angle observed in the experiment varies, from a value larger than $85^\circ$ prior to the critical point, to a value as low as $60^\circ$ during the acceleration phase. Furthermore, these angles appear to depend on the syringe speed, $U$, and have a strong effect on the resultant dynamics~\cite{qian2008}.  A second reason for the degraded agreement between experiment and simulation may be due to high radial velocities observed which violate the assumptions in the the one-dimensional numerical approach considered here, although despite this, the numerical predictions are suprisingly faithful.

In summary, the retraction speed of the syringe can exert a huge influence on the size of the resulting droplet that remains on the substrate, with the transition between the three regimes identified being determined by the balance between the flow into the liquid bridge and the onset of the pinch-off instability. At the highest retraction speeds, the small droplet size appears to be determined by the contact line speed, raising the possibility that even smaller droplets might be achieved on surfaces that are smoother and/or exhibit higher contact angles.  The numerical model provides surprisingly accurate predictions of the dynamices, even in the regimes where the contact line motion and the presence of viscous and inertial forces make the one-dimensional assumptions questionable.

\noindent The work was supported by NSF/Sandia cooperative research agreement and by Nordson/EFD Corporation.

\end{document}